\documentclass[12pt]{article}
\usepackage{osa2}
\usepackage{overcite}

\begin{document}
\title{Extinction theorem for ultrafast pulses} 

\author{Cs. Benedek, M. G. Benedict and T. Ser\'enyi}
\address{Department of Theoretical Physics, University of Szeged,
Tisza krt 84., H-6720, Szeged, Hungary}
\email{benedict@physx.u-szeged.hu}


\begin{abstract}
{Instead of using a frequency dependent refractive index, 
we propose to use the extinction theorem to describe 
reflection and transmission of an ultrashort 
pulse passing through the boundary. 
When the duration of the pulse is comparable 
with the relaxation time, the results differ significantly 
from those given by the traditional method, especially if 
the carrier frequency is close to an absorbtion line. 
We compare the two approaches using the data of GaAs in
the infrared domain.}
\end{abstract}
\ocis{320.7120, 350.5500}



\noindent
It is well known that the reflection and refraction of an electromagnetic
wave on the boundary of a material system is a light scattering phenomenon.
The atoms in the medium are brought into oscillations by the incoming wave,
and then secondary waves are generated by the induced atomic dipoles. The
transmitted wave is the result of superposition of the scattered wave and
the external field, while the reflected wave is a result of backscattering.
One side of the process, the solution of the quantum mechanical equations of
motion of the perturbed atoms (or the forced oscillations in the classical
model) is treated in most textbooks in order to calculate the frequency
dependence of the polarizability and the dielectric constant. 
The other side of the problem, the actual scattering process is usually
replaced by phenomenological boundary conditions  -- Fresnel formulas --
for the macroscopic fields.
One then uses a frequency dependent index of refraction, and calculates the
response for each spectral component. 

Instead of that traditional approach
which neglects the transient regime of the forced oscillations, we use a
procedure that exhibits the actual physical process. This is the
method of integral equations, known also as the Ewald-Oseen extinction
theorem in the stationary situation~\cite{BWPO}. 
We are going to consider here the one-dimensional variant of this theory
applied to ultrafast pulses.

We note that local field effects influence the coupled atom-field
dynamics, therefore it can affect the reflection and transmission properties
too. For weak fields this leads only to a constant shift \cite{SRAD} of the
resonance frequencies, therefore we shall treat the problem here without this
correction.

The extinction theorem has been applied for the resonant 
nonlinear case in our works \cite{SRAD,BT88,BMTZ91} 
but with the slowly
varying amplitude approximation (SVEA) in time. 
The first model calculation for a linear medium 
treating the full time development of a pulse without SVEA  has been given
by one of us \cite{spie}.


We consider the boundary
value problem for the transmission of a light pulse through a plane-parallel
resonant layer consisting of polarizable atoms. Let the incoming field be a
linearly polarized plane wave: 
\begin{equation}
E_{in}(x,t)=E(t-x/c)  \label{e1}
\end{equation}
We restrict  ourselves here to the case of normal incidence. 
Then the problem is one dimensional in space, thus the macroscopic field obeys
the inhomogeneous wave equation: 
\begin{equation}
{\frac{\partial ^{2}E}{\partial x^{2}}}-{\frac{1}{c^{2}}}{\frac{\partial
^{2}E}{\partial t^{2}}}={\frac{1}{\varepsilon _{0}c^{2}}}{\frac{\partial ^{2}%
{P}}{{\partial t^{2}}}}.  \label{waveq}
\end{equation}
In the situation
considered here, ${P}$ is different from zero in a slab placed between
$x=0$ and $x=L$, and the solution of Eq.(\ref{waveq}) has the form\cite{SRAD}: 
\begin{equation}
E(x,t)=E_{in}(x,t)-{\frac{1}{2\varepsilon _{0}c}}{\int_{0}^{L}}{{\frac{%
\partial {P}}{{\partial t}}}\ (x^{\prime },{t-|x-x^{\prime }|}/c)}%
dx^{\prime }.   \label{inteq}
\end{equation}
Here $E_{in}$ is the solution of the homogeneous equation corresponding to (%
\ref{waveq}), and it is identified with the incoming wave, while the
second term
is a scattered wave which is the superposition of outgoing elementary waves
originating in different $x^{\prime }$ planes. 
Given an incoming wave, the transmitted wave is determined by the whole
solution (\ref{inteq}) at $x\geq L$, while the reflected wave is described
only by the second, integral term at $x\leq 0$. 

As we are considering the linear case, the dynamics of the polarization $%
{P}$ in the medium can be determined by a first order perturbation
theory. Supposing that at the beginning the system is in its ground state,
the dipole moment density induced by the field is given by the following
expression: 
\begin{eqnarray}
{P}(t,x) = \sum_j {P}_j = 
2N Re\left[\sum_{j} {\frac{i|d_{j}|^{2}}{\hbar }}e^{(-i\omega _{j}-%
\frac{\gamma _{j}}{2})t}  \right. \times  
\left. \int_{0}^{t}E(t^{\prime },x)e^{(i\omega _{j}+\frac{\gamma _{j}}{2}%
)t^{\prime }}dt^{\prime }\right]  \label{pert}
\end{eqnarray}
where $N$ is the number density of the dipoles in the medium, $d_{j}$ is the
transition dipole matrix element between the ground state and the $j$-th
excited level, 
$\omega _{j}$ and $\gamma _{j}$
are the corresponding transition frequencies and decay constants, and $%
E(t,x) $ is the field strength at the position $x$ of the atomic dipole.
In the ordinary theory of dispersion the time dependence of the electric
field, $E(t)$ is taken to be $E_{0}e^{-i\omega t}$, and after performing the 
integration in (\ref{pert}) 
the terms containing $e^{-\gamma t/2}$ are omitted, as they are rapidly
decaying. In this way one obtains a frequency dependent 
susceptibility 
$\chi (\omega )=(1/\varepsilon_{0})\sum_{j}( 2N{|d_{j}|^{2}/\hbar}) /(\omega
_{j}-\omega -i\gamma_{j}/2), \label{szusz}
$
%
and refractive index $n=(1+\chi(\omega))^{1/2}$.  
%
%
Our main point is that if the duration of the 
whole pulse is shorter than the decay time of the
terms which contain $e^{-\gamma_{j} t/2}$, this traditional 
approach fails, and all the terms following from (\ref{pert}) should be kept.


In the case of an optically thin layer, the calculations can be simplified.
Then $L/\lambda \ll 1$, the spatial
variation of the polarization can be neglected, and instead of integrating
in Eq.(\ref{inteq}) we can use the mean value of the time derivative of the
polarization. 
%
In this thin medium case 
one can derive a system of 
coupled differential equations for the polarization components, ${P}_{j}:$%
\begin{eqnarray}
\stackrel{..}{{P}_{j}}+\stackrel{.}{{P}_{j}}{\gamma
_{j}}+{P}_{j}(\frac{\gamma _{j}^{2}}{4}+{\omega
_{j}^{2}}) &=&  
2N\frac{|d_{j}|^{2}}{\hbar }(\omega _{j}E_{in}-\frac{L}{2\varepsilon {_{0}}%
c}\omega _{j}\sum_{k}\stackrel{.}{{P}_{k}})  \label{diffegy}
\end{eqnarray}
For an arbitrary thick layer 
one has to solve the coupled
integro-differential equations (\ref{inteq}) and (\ref{pert}), 
with a numerical method.


We have performed calculations, based on the theory given
here and compared the results with those of the traditional 
method of Fresnel formulas. 
For definitness we have used the data\cite{Palik} of GaAs in the infrared domain.
Two resonant frequencies and the 
with constants 
%
%
$\lambda_1=37.31\mu m$, $\lambda_2=0.4305\mu m$,
$\gamma_1=0.034\omega_1$, $\gamma_2=0$, ($\lambda_i=2\pi c/\omega_i$),
and 
${\displaystyle{s_i:=\frac{2\ N|d_i|^2}{\varepsilon_0\hbar\omega_i}}}$,
with $s_1=9.89$,  $s_2=2.07$. 
reproduced very well the measured\cite{Palik}
stationary 
dispersion and  absorbtion of the material in the range 
$\lambda= 1 - 10^2\mu m$. 
%
%
%
%
%
For the sake of simplicity we have chosen the following form of the incoming
pulse $E_{in}(t)=\sin ^{2}\Omega t\cos\omega_{0} t$, $0<t<\pi/\Omega$.  
As we are considering the ultrafast regime we have chosen 
$\Omega=0.25 \omega_{0}$. The shape of this incoming pulse is
shown in Fig. 1(a) below.
%
The solution of Eq.(\ref{diffegy}) corresponding to this incoming pulse 
and to the initial conditions 
${P}_{j}(0)=0$, $\dot{P}_{j}(0)=0$, $j=1,2$
can be given for the thin layer by an exact but lengthy formula.   


Far from resonance $(\omega_{0} \ll \omega_{1}$, $\omega_{0} \gg \omega_{1})$
both methods predict the same results.
On the other hand significant differences can be 
seen for pulses with $ \omega_{0} \approx \omega_{1} $, 
i.e. close to resonance, but still in the linear regime. 
The calculations were performed with $ \omega_{0} = 1.03 \omega_{1} $, 
for layers of different thicknesses of the order 
of the central wavelength, $\lambda_0=2\pi c/\omega_0$.
Fig. 1 shows the results obtained 
for an optically thin layer $L=\lambda_{0}/256$. In that case we have solved 
the system of equations (\ref{diffegy}).
The numerical solution of Eqs. (\ref{inteq}) and (\ref{pert}) 
are shown in Fig. 2 for a layer of thickness of $\lambda_0$.
The continuous lines are the results obtained from the present 
transient extinction theory, while the dashed lines  are
the amplitudes obtained by using the frequency dependent 
index of refraction. 
It can be seen, that some of the overall characteristics of the response obtained by 
the two different methods are similar.  Both treatments predict significant broadening
of the reflected and transmitted pulses, since the decay time, $2/\gamma$ is long 
compared with the duration of the excitation. 
In addition, till the end of the exciting pulse, $\omega_0 t=4 \pi$, 
the reflected amplitudes are very similar
except for a little phase shift. The transmitted pulse appears in both cases  
with the same expected delay that can be calculated as $t_g=v_g/L$,
where $v_g$ is the the group velocity  \cite{BB,OS}.  
This gives $\omega_0 t_g=7.96\pi$ in good agreement with the numerical 
result.
We note, in addition, that both methods yielded  precursors \cite{BB,OS} 
in the transmitted fields appearing at $t=L/c$,  not seen in the figures, 
because of their smallness. 

In the case of the thin layer the forms of the fields are quite similar, see Fig 1.
On the other hand significant differences are seen in the time dependences 
of the transmitted and reflected pulses calculated by the two different treatments in
layers of thickness comparable with $\lambda_{0}$. 
We attribute these differences in the amplitudes,
as well as in the phases to the approximative 
character of the time dependence of the polarization in the Fourier 
method, i.e. to the omission of the transients at the beginning and at the end of the
excitation. This part is non-negligible if the duration of the process
is comparable with the relaxation time since it represents an important contribution 
to the elementary waves generated by the induced dipoles.

In conclusion, we find that for ultrafast pulses the properties 
of a dispersive medium  must be calculated by taking care of the transient
response of the system.  This can be done best by using a fully
time dependent treatment as proposed in the present work.  

We thank Z. Bor and Z. Horv\'ath for stimulating discussions. 
This work was supported by the Hungarian Scientific Research Fund 
(OTKA) under contracts T022281 and T32920.   
 

\newpage
\section*{List of figures}

\vskip.5in

\begin{figure}[h]
\caption{Incoming (a),  reflected (b) and transmitted (c) amplitudes 
for a layer of thickness $L=\lambda/256$. In (b) and (c) 
continuous lines are calculated from the extinction theorem, 
while he dashed lines are obtained using the index of refraction.}
\end{figure}

\begin{figure}
\caption{Reflected (a) and transmitted (b) pulses calculated 
by the two different methods (see caption of Fig 1.) 
for the incident pulse shown in Fig. 1 (a), 
and for a layer of thickness $ L=\lambda_{0} $, 
In  (c) the origin has been shifted to $\omega_0t =2\pi$ corresponding to 
$t=L/c$ (the transit time in vacuum)} 
\end{figure}

\newpage

\centerline{\scalebox{1}{\includegraphics[9cm,7cm]{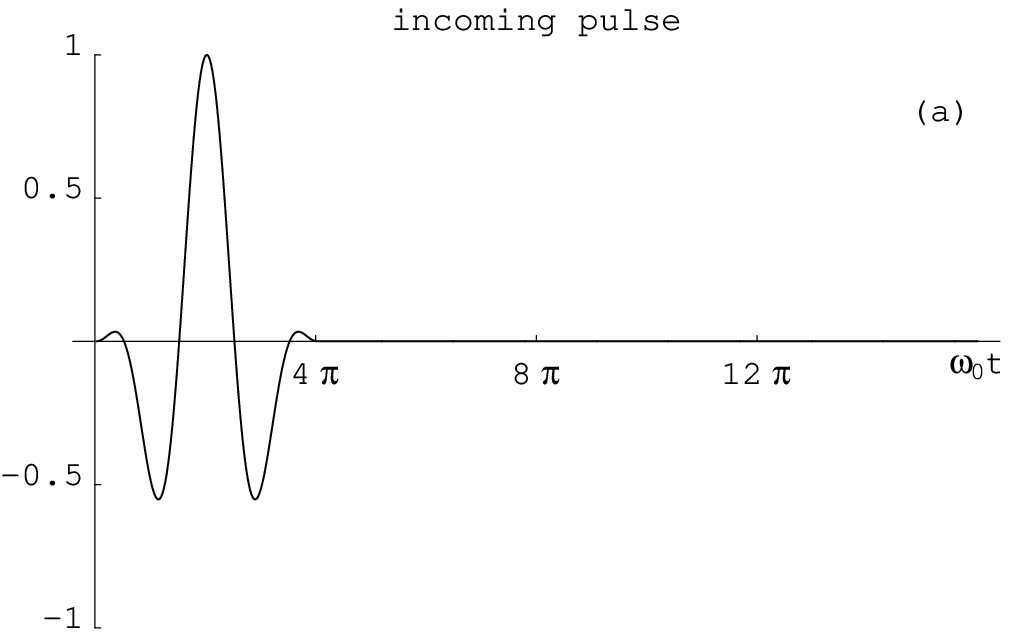}}}
\centerline{\scalebox{1}{\includegraphics[9cm,7cm]{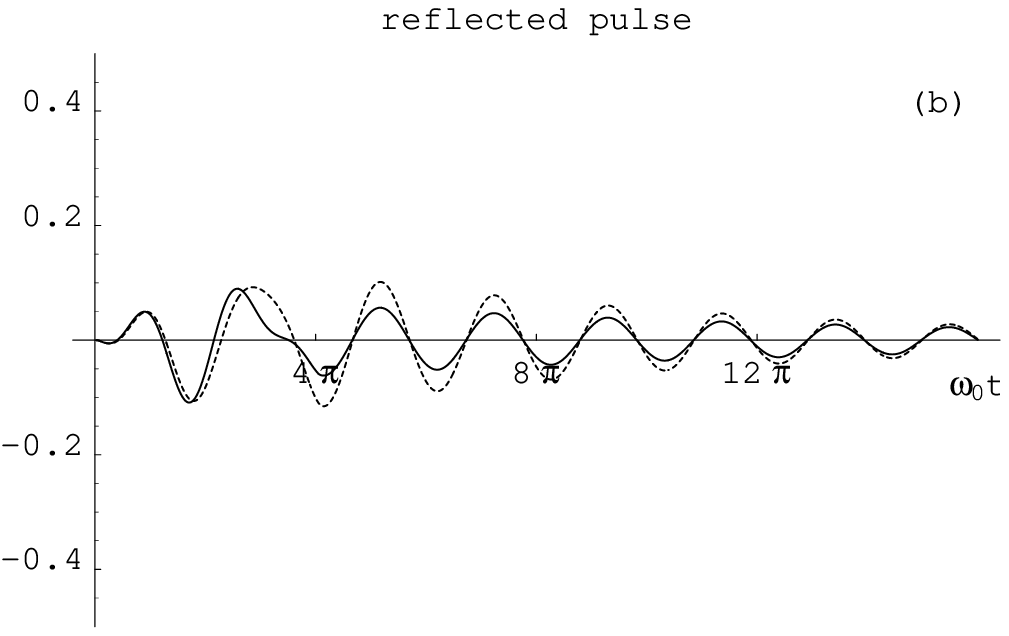}}}
\centerline{\scalebox{1}{\includegraphics[9cm,7cm]{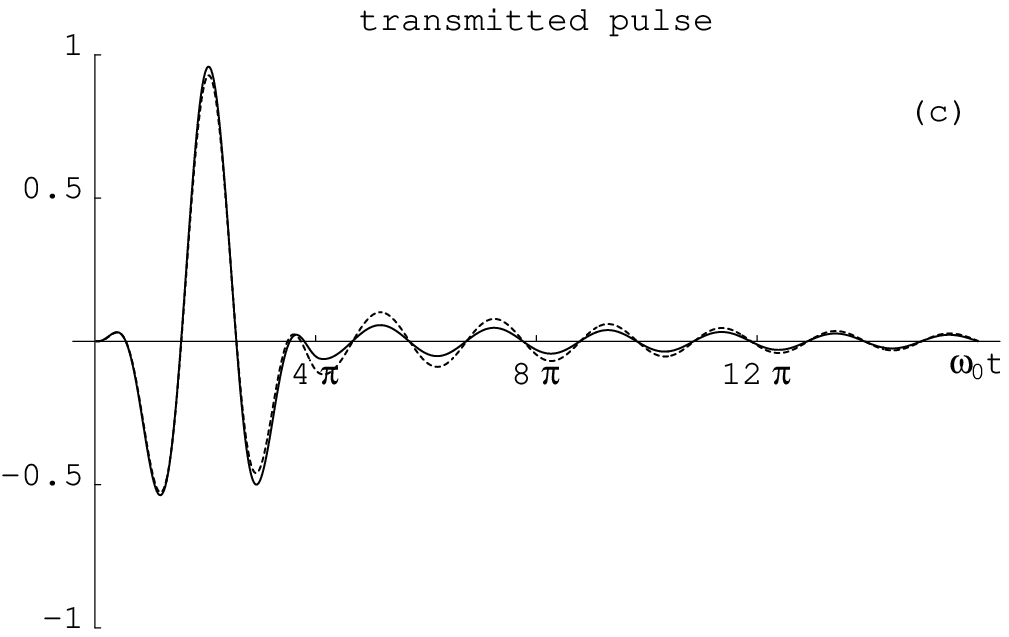}}}
Figure 1. Cs. Benedek, Optics Letters

\newpage
\centerline{\scalebox{1}{\includegraphics[9cm,7cm]{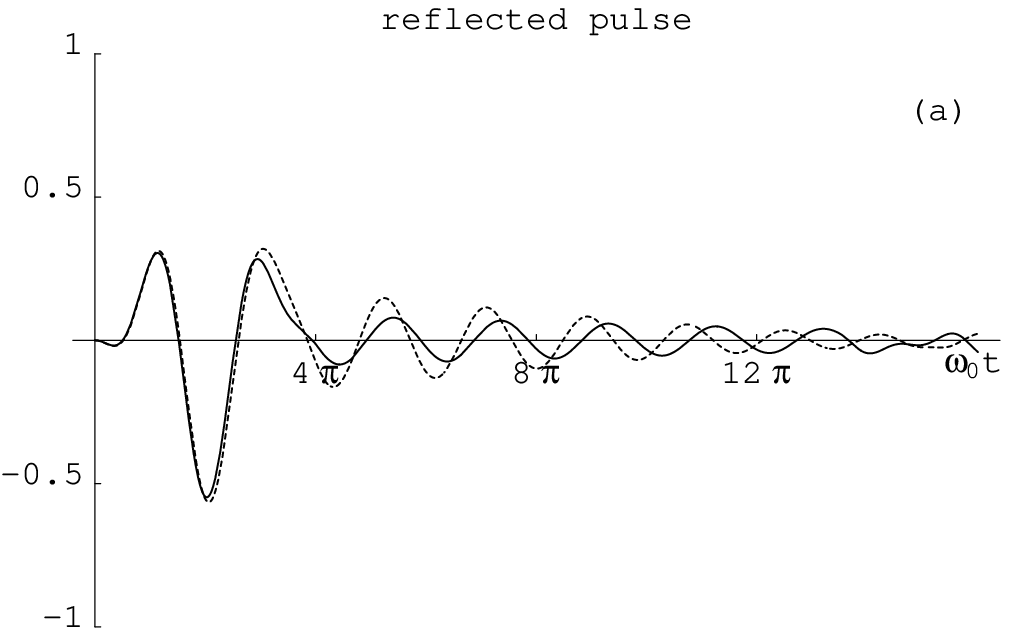}}}
\centerline{\scalebox{1}{\includegraphics[9cm,7cm]{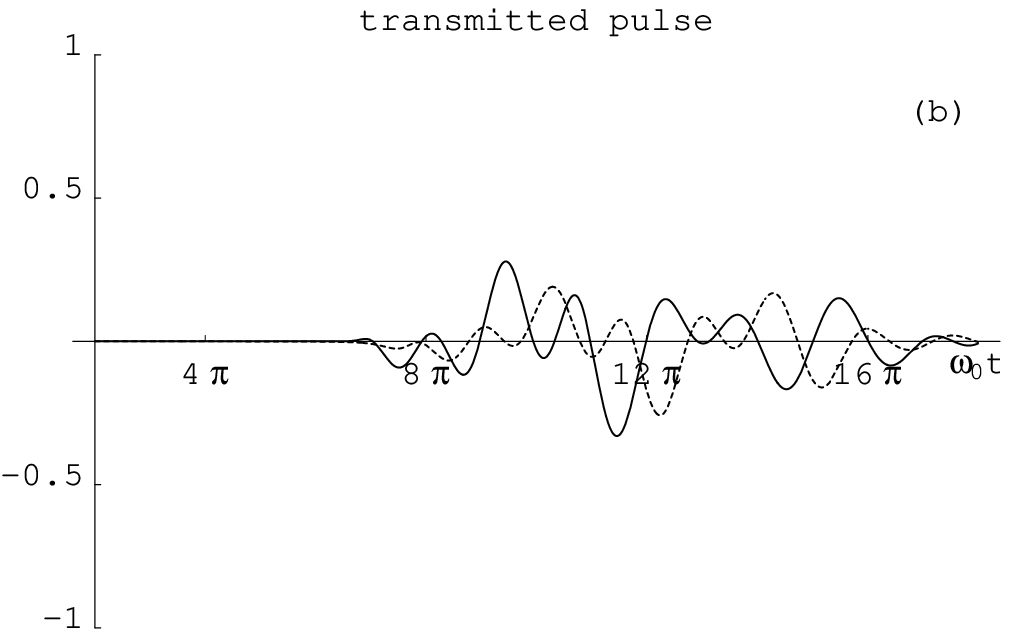}}}
\vskip3in
Figure 2. Cs. Benedek, Optics Letters


\begin{thebibliography}{9}
\bibitem{BWPO}  M. Born and E. Wolf, {\it Principles of optics} 6th ed.
(Pergamon, Oxford, 1989).
%
\bibitem{SRAD}  M.G. Benedict, A.M. Ermolaev, V.A. Malyshev, I.V. Sokolov,
and E.D. Trifonov, {\it Superradiance} (IOP, Bristol, 1996).
%
\bibitem{BT88}  M.G. Benedict, E.D. Trifonov, ``Coherent reflection as
superradiation on the boundary of a resonant medium",  Phys. Rev. A 
{\bf 38}, 2854-2862 (1988).
%
\bibitem{BMTZ91}  M.G. Benedict, V. A. Malyshev, E.D. Trifonov, A. I Zaitsev
``Reflection and transmission of ultrashort light pulses through a thin
resonant medium: Local field effects"  Phys. Rev. A {\bf 43},
3845-3853 (1991).
%
\bibitem{Palik} E. D. Palik, "Gallium Arsenide" in {\it Handbook of Optical Constants of Solids } E. D.
Palik, ed. (Academic Press, 1985), pp. 429-443.
%
\bibitem{spie}  M.G. Benedict, 
"On the reflection and transmission of femtosecond pulses", 
Proceedings of SPIE {\bf 3573} 486-489 (1998). 
%
\bibitem{BB} L. Brillouin, {\it Wave propagation and group velocity} (Academic Press, New York and London, 1960).
%
\bibitem{OS} K. E. Oughstun, O. C. Sherman, {\it Electromagnetic pulse propagation in causal dielectrics} (Springer-Verlag, Berlin, 1994).
\end{thebibliography}
\end{document}